# CMOS Monolithic Active Pixel Sensors (MAPS) for future vertex detectors

R. Turchetta
*CCLRC, Rutherford Appleton Laboratory, Chilton, Didcot, Oxfordshire, OX11 0QX, UK*

This paper reviews the development of CMOS Monolithic Active Pixel Sensors (MAPS) for future vertex detectors. MAPS are developed in a standard CMOS technology. In the imaging field, where the technology found its first applications, they are also known as CMOS Image Sensors. The use of MAPS as a detector for particle physics was first proposed at the end of 1999. Since then, their good performance in terms of spatial resolution, efficiency, radiation hardness have been demonstrated and work is now well under way to deliver the first MAPS-based vertex detectors.

## 1. INTRODUCTION

In the early '90s, CMOS Monolithic Active Pixel Sensors were invented for the detection of visible light. It was immediately recognised they promise several advantages [1, 2] over existing imaging devices, in terms of functionality, power, radiation hardness, speed and ease of use. However, at the beginning the relatively poor performance of CMOS sensors limited their use to some specific applications, like web cams, toy cameras, etc.. Continuous improvement in the technology, reducing the dark current and improving the electronic noise, made CMOS sensors gain more and more ground and today, CMOS sensors are becoming the dominant image sensing device. Research in this area is thriving and CMOS sensors are widely used, including in areas where low noise performance is required, as in commercial digital cameras or in scientific applications [3, 4, 5, 6].

For particle physics, the concept underlying the use of CMOS sensors for the detection of charged particles was proposed [7] a few years ago and demonstrated on small prototypes. In order to obtain a sensor usable in a particle physics experiment, key performance aspects like radiation hardness and low noise need to be demonstrated for a large area sensor. This paper reviews some of the main results obtained so far and the R&D work undergoing in several laboratories to develop the first MAPS-based vertex detectors. The paper will conclude on a brief review of technology development on CMOS technology for image sensors.

## 2. PERFORMANCE

### 2.1. Baseline 3T pixel

In its simplest form, the pixel architecture of a MAPS features three transistors (fig. 1): transistor MR acts as a switch to reset the floating node FD on the photodiode PD; MS also acts a switch to select the pixel for readout and MI is the input of a source follower whose current source is located outside the pixel and is common to all the pixel in a column. Different types of diodes are available in a standard CMOS process, but for efficient detection of charged particles N-well diodes to the P-substrate are used. This structure was first proposed to increase the quantum efficiency of imaging sensors [2], and then proposed as a 100% efficient sensor for particle physics [7]. In many technologies, a deep N-well to P-substrate diode is also available. This diode could provide faster collection time and hence better radiation hardness [8, 9]. In most technologies, the substrate is a relatively low doped, ~10 Ohm cm resistivity, epitaxial layer, created on top of a thick, hundreds of microns, heavily doped P substrate. The thickness of the epitaxial layer is normally in the range of a few microns up to 20.

This 3T pixel was widely used in the first designs of a sensor for particle physics [10, 11, 12, 13, 14]. This sensor works in the rolling shutter mode: all the pixels in one row are read out in parallel by selecting the MS switch and then reset by selecting the MR switch, which resets the FD node to a voltage close to VRST. This architecture was tested on a number of technology, with feature size ranging from 0.18 up to 0.6 microns and epitaxial layer thickness ranging from 2 up to 20. In some sensors, the epitaxial layer was absent and the substrate was low doped. The sensors work also in this case although the absence of a sensitive volume definition can lead to a higher spread of charge between pixels.

In all these structures, charge collection is mainly by diffusion, and then a relatively slow process, although deep n-well structures [8, 15]can speed up the process. As a result, radiation hardness is limited by the lifetime of minority carriers. By carefully choosing the pixel architecture, sensors have been found to perform with S/N greater than 10 for a 8 micron epitaxial layer and up to $10^{14}$ p/cm$^2$ [13].

### 2.2. Advanced pixels

Although the 3T is a well tested and known architecture, its speed is limited by the rolling shutter operation. This is alright for some specific applications, e.. the Star vertex detector at RHIC. For the International Linear Collider ILC or for the B-factory at KEK, much higher speed is demanded. Pipeline architectures have been proposed.

For the ILC, the so-called Flexible Active Pixel Sensor architecture was proposed [16, 17]. In the pixel, 10 memory cells are located. Data are written at high speed in the cells and are retrieved later at a slower speed, taking advantage of the beam structure, which at ILC features a 199-ms long beam off periods for every ms period of beam on.

A similar architecture, called the Continuous Acquisition Pixel (CAP) is also proposed for the B-factory at KEK [18]. More details can be found in these proceedings [19].

Data sparsification has also been developed and good results were presented in [20]. It is worth mentioning that the development of a digital MAPS has also started for the electromagnetic calorimeter Calice at ILC [21].

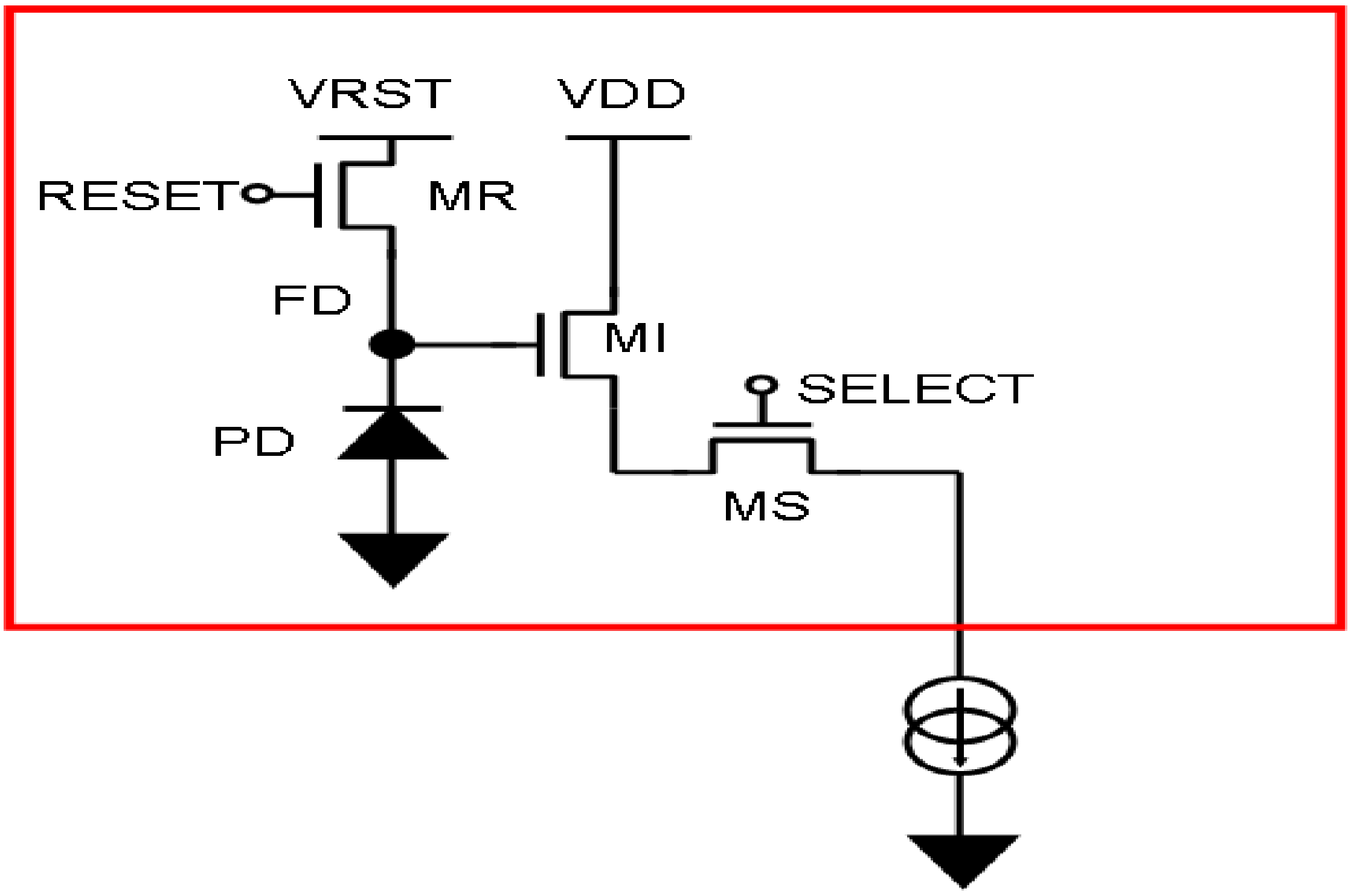

**Figure 1. Schematic of a 3T Pixel**

## 3. CONCLUSION. CMOS TECHNOLOGY FOR IMAGE SENSORS

Active Pixel Sensors for imaging were first produced in the early '90s. At that time, the performance of the sensors in terms of noise and leakage current were somehow limited. However, during the years, there have been a number of development which has greatly improved their performance. This improvement are motivated by the need of better sensors for today's imaging devices, whose demand is continuously growing. Cameras are now added to many consumer devices: mobile phones, toys, web cams, computers, … Among the main developments, two are worth mentioning because of interest also for particle physics: the 4T pixel with pinned diode and stitching.

A schematic of a 4T pixel with pinned photodiode is shown in figure 2. With respect to the 3T architecture, two are the major changes. A fourth transistor MX is added in series between the floating node FD and the photodiode pPD. With proper timing of the gate TX, the transistor MX can transfer the charge from the pinned photodiode pPD to the floating node, allowing true in-pixel Correlated Double Sampling (CDS), thus eliminating the reset noise, which would otherwise be the main noise source. The pinned photodiode has the same structure of a floating base bipolar transistor, as an additional p+ implant is added on top of the N-well/p substrate structure (see for example [22]). With a proper doping profile, the pinned photodiode can be made such that full charge transfer is obtained when the transfer gate MX is operated. This pinned photodiode is today widely used in cameras, as it provides low noise operations. However, the operation of the diode limits its voltage swing [23]. This should not be a problem for particle physics experiment. It is worth mentioning that the doping profiles of the pinned photodiode-transfer gate structure are highly optimized and so it would be interesting to assess its radiation hardness. At present, no result has been published on this aspect.

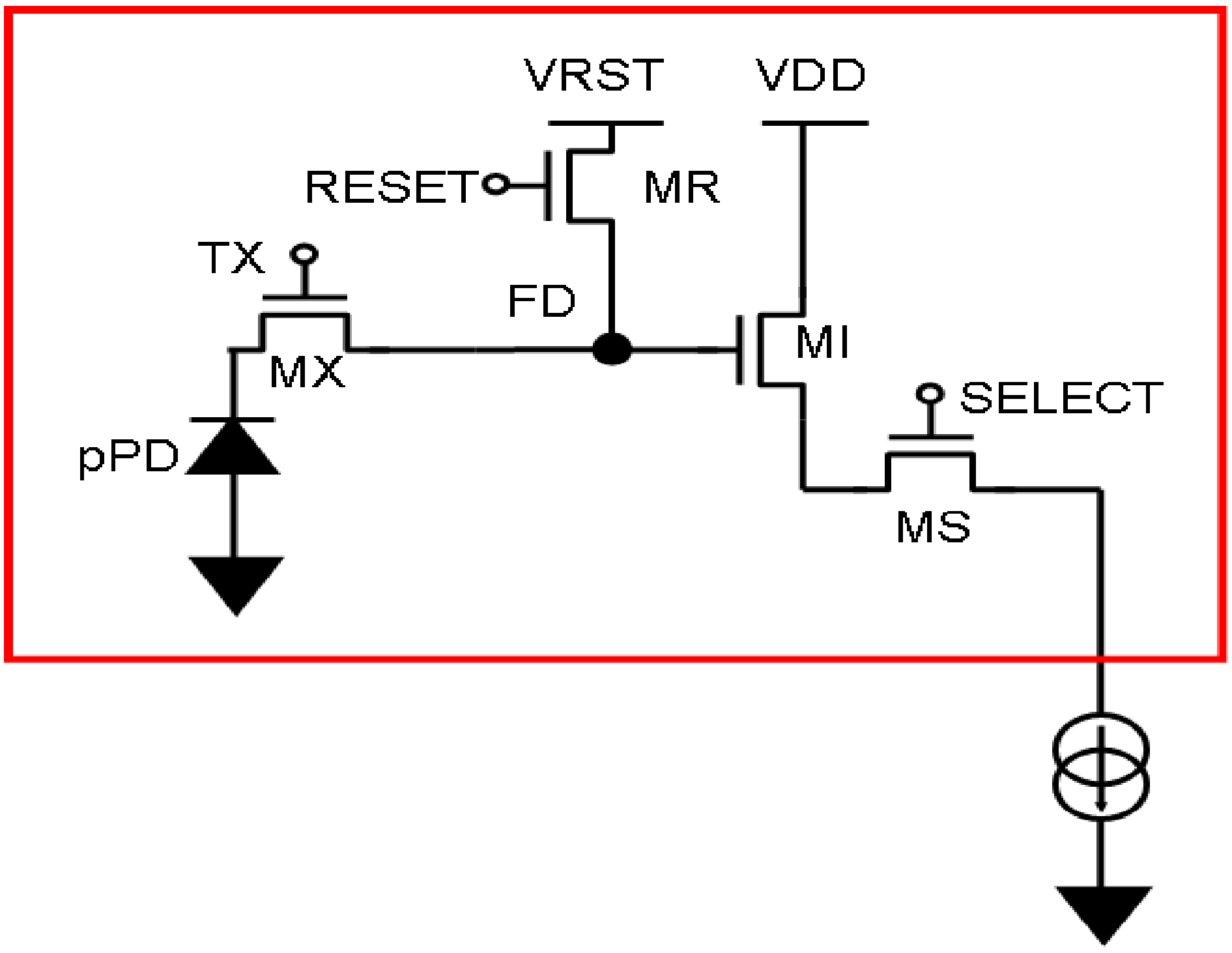

**Figure 2. Schematic of a 4T Pixel**

Stitching is a well-known technique, which allow production of large, monolithic structures. In CMOS foundries, the size of a circuit is normally limited by the size of the reticle, i.e. the largest areas that can be exposed in one shot. This area is normally of the size of about 2cmx2cm, although slightly larger reticles can be found in the industry. In a normal production, reticles are stepped and repeated across the wafer at a step determined by the size of the reticle. However, it is possible to obtained seamless structures that are larger than the reticles by carefully optimizing the step and repeat operation. This technique, called stitching, is well known and has been used in the industry for many years, for example in order to produce large areas charge-coupled devices. Today, driven by the manufacturing of image sensors, which would replace the 35 mm film, many CMOS foundries propose stitching as a standard foundry service (see for example [24]). Generally speaking, a sensor can be made as large as the wafer, although yield consideration might limit the size of the sensor one wants to make. Stitched sensors can already be found in consumer cameras, like for example [25].

Since they were first proposed for their use as particle detectors a few years ago, CMOS sensors have been demonstrated and today we are close to see the first MAPS-based vertex detectors. Continuous advances in the image sensors technology make these devices even more appealing and, as technology progresses, we are likely to see more and more experiments adopting this technology

## Acknowledgments

The author wish to thank all the people who contributed to the development of MAPS for particle physics in RAL, J. Crooks, L. Jones, M.Tyndel, G. Villani, F. Zakopoulos, in the University of Liverpool, P.P. Allport, G. Casse, J. Velthuis[1], in the University of Glasgow, R. Bates, V. O'Shea, in the Imperial College London, P. Dauncey, and in the University of Birmingham, N. Watson.

The development of MAPS for particle physics was funded by PPARC, under a PPRP grant.

## References

[1] B. Pain, S. K. Mendis, R. C. Schober, R. H. Nixon, E. R. Fossum, *Low-power, low-noise analog circuits for on-focal-plane signal processing of infrared sensors*, Proceedings of SPIE, vol. 1946, 1993, 365-374

[2] B. Dierickx, G. Meynants, D. Scheffer, *Near 100% fill factor CMOS active pixels*, 1997 IEEE CCD & Advanced Image Sensors Workshop, Brugge, Belgium

[3] N. Waltham et al., presented at PSD7, Liverpool, September, 2005, to be published in Nucl. Instruments and Methods A

[4] M.L.Prydderch, N.J.Waltham, R.Turchetta, M.J.French, R.Holt, A.Marshall, D.Burt, R.Bell, P.Pool, C.Eyles, H.Mapson-Menard, *A 512x512CMOS*

[1] Present address: Bonn University, Germany

*Monolithic Active Pixel Sensor with Integrated ADCs for Space Science*, Nucl. Instr. and Meth. A, vol. 512, no. 1-2, page 358-367

[5] Q.R. Morrissey, N.R. Waltham, R. Turchetta, M.J. French, D.M. Bagnall, B.M. Al-Hashimi, *Design of a 3μm pixel linear CMOS sensor for Earth observation*, Nucl. Instr. and Methods A, vol. 512, no. 1-2, page 350-357

[6] A. Fant et al., I-IMAS: a 1.5D sensor for high resolution scanning, presented at PSD7, Liverpool, September, 2005, to be published in Nucl. Instruments and Methods A

[7] R. Turchetta et al., A monolithic active pixel sensor for charged particle tracking and imaging using standard VLSI CMOS technology, Nuclear Instruments and Methods A 458 (2001) 677-689

[8] G. Villani, R. Turchetta and M. Tyndel, "Analysis and simulation of charge collection in Monolithic Active Pixel Sensors (MAPS)", presented at the 8th Topical Seminar on Innovative Particle and Radiation Detectors, 21-24 October 2002, Siena (Italy), submitted to Nucl. Phys. B

[9] G.Villani, P. P. Allport, G. Casse, A. Evans, R. Turchetta, J. J. Velthuis, *Design and characterization of a novel, radiation- resistant active pixel sensor in a standard 0.25 μm CMOS technology*, proceedings of the IEEE/NSS conference, Rome, Italy, 2004 to be published in Trans. Nucl Sci.

[10] G. Deptuch et al., *Simulation and Measurements of Charge Collection in Monolithic Active Pixel Sensors*, presented at "Pixel2000, International Workshop on Semiconductor Pixel Detectors for Particles and X-Rays", Genova (Italy), 5-8 June 2000, to be published on NIM A

[11] G. Claus et al., *Particle Tracking Using CMOS Monolithic Active Pixel Sensor*, presented at "Pixel2000, International Workshop on Semiconductor Pixel Detectors for Particles and X-Rays", Genova (Italy), 5-8 June 2000, to be published on *NIM A*

[12] G. Deptuch et al., *Design and Testing of Monolithic Active Pixel Sensors for Charged Particle Tracking*, presented at IEEE Nucl. Science Symposium and Medical Imaging Conference, Lyon (France), 15-20 October 2000, submitted to IEEE Trans. on Nucl. Science

[13] J. J. Velthuis, P. P. Allport, G. Casse, A. Evans, R. Turchetta, G. Villani, *CMOS sensors in 0.25 μm CMOS technology*, proceedings of Vertex 2004 to be published in Nucl. Inst. Meth. A.

[14] J. J. Velthuis, P. P. Allport, G. Casse, A. Evans, R. Turchetta, G. Villani, *Characterization of active pixel sensors in 0.25μm CMOS*, proceedings of the IEEE/NSS conference, Rome, Italy, 2004 to be published in Trans. Nucl Sci.

[15] L. Ratti et al, these proceedings

[16] R. Turchetta et al., *FAPS, a CMOS sensor with multiple storage for fast imaging scientific applications*, Extended programme of the 2003 IEEE Workshop on CCD and Advanced Image Sensors

[17] P.P. Allport, G. Casse, A. Evans, L. Jones, R.Turchetta, M.Tyndel, J.J. Velthuis, G. Villani, F. Zakopoulos, *CMOS Sensors for High Energy Physics*, Extended programme of the 2005 IEEE Workshop on CCD and Advanced Image Sensors

[18] G. Varner et al., *Development of the Continuous Acquisition Pixel (CAP) sensor for High Luminosity Lepton Colliders*, presented at IEEE Nucl. Science Symposium and Medical Imaging Conference, October 2005, submitted to IEEE Trans. on Nucl. Science

[19] G. Varner et al., these proceedings

[20] N. Fourches et al., *Performance of a Fast Programmable Active Pixel Sensor Chip Designed for Charged Particle Detection*, presented at IEEE Nucl. Science Symposium and Medical Imaging Conference, October 2005, submitted to IEEE Trans. on Nucl. Science

[21] http://www.hep.ph.ic.ac.uk/calice

[22] R. M. Guidash et al., *A 0.6 μm CMOS pinned photodiode color imager technology*, in IEDM Tech. Dig., 1997, p. 927-928

[23] T. Lule et al., *Sensitivity of CMOS based imagers and scaling*, IEEE Trans. on Electron Devices, vol. 47, n. 11, p. 2110-2122, (2000)

[24] http://www.towersemi.com

[25] http://www.canon.com/technology/d35mm/02.html